\documentclass[reprint, amsmath, amssymb, aps, prd, nofootinbib, superscriptaddress]{revtex4-1}

\usepackage{orcidlink} 

\def\checkmark{\tikz\fill[scale=0.4](0,.35) -- (.25,0) -- (1,.7) -- (.25,.15) -- cycle;} 
\newcommand{\crossmark}{$\mathbin{\tikz [x=1.4ex,y=1.4ex,line width=.2ex] \draw (0,0) -- (1,1) (0,1) -- (1,0);}$} 
\newcommand{\Yorsh}{\emph{Yorsh}}
\newcommand{\Sangria}{\emph{Sangria}}

\def\mnras{\rm{MNRAS}}
\def\prd{\rm{PRD}}
\def\prl{\rm{PRL}}

\newcommand{\bham}{Institute for Gravitational Wave Astronomy \& School of Physics and Astronomy, University of Birmingham, Edgbaston, Birmingham B15 2TT, UK}
\newcommand{\IoA}{Institute of Astronomy, University of Cambridge, Madingley Road, Cambridge, CB3 0HA, UK}
\newcommand{\KICC}{Kavli Institute for Cosmology, University of Cambridge, Madingley Road, Cambridge, CB3 0HA, UK}
\newcommand{\DAMTP}{Department of Applied Mathematics and Theoretical Physics, Centre for Mathematical Sciences, University of Cambridge, Wilberforce Road, Cambridge, CB3 0WA, UK}

\begin{document}

\title{Searching for stellar-origin binary black holes in LISA Data Challenge 1b: Yorsh}

\author{Diganta Bandopadhyay
\orcidlink{0000-0003-0975-5613}}
\email{diganta@star.sr.bham.ac.uk}
\affiliation{\bham}

\author{Christopher J.\ Moore
\orcidlink{0000-0002-2527-0213}}
\email{cjm96@cam.ac.uk}
\affiliation{\IoA}
\affiliation{\DAMTP}
\affiliation{\KICC}

\date{\today}

\begin{abstract}
    This paper reports the first search for stellar-origin binary black holes within the LISA Data Challenges (LDC).
    The search algorithm and the \Yorsh{} LDC datasets, both previously described elsewhere, are only summarized briefly; the primary focus here is to present the results of applying the search to the challenge of data.
    The search employs a hierarchical approach, leveraging semi-coherent matching of template waveforms to the data using a variable number of segments, combined with a particle swarm algorithm for parameter space exploration.
    The computational pipeline is accelerated using graphical processing unit (GPU) hardware.
    The results of two searches using different models of the LISA response are presented.
    The most effective search finds all five sources in the data challenge with injected signal-to-noise ratios $\gtrsim 12$. 
    Rapid parameter estimation is performed for these sources.
\end{abstract}

\maketitle

\section{Introduction}\label{sec:introduction}

The Laser Interferometer Space Antenna (LISA) will detect gravitational waves (GWs) in the millihertz frequency band \cite{2024arXiv240207571C}.
Stellar-origin binary black holes (SoBBHs) with orbital frequencies in this bandwidth \cite{2016MNRAS.462.2177K, 2016PhRvL.116w1102S} are the progenitors of the black hole merger signals observed by LIGO/Virgo \cite{2023PhRvX..13d1039A}.  
LISA is expected to be able to detect a small number of these systems \cite{2022MNRAS.514.4669S, 2019PhRvD..99j3004G}.

SoBBHs in the LISA band produce long-lived, broadband GW signals that pose significant data-analysis challenges.
If these signals can be detected, their multiyear durations allow certain parameter combinations (such as chirp mass, time to merger, and eccentricity) to be extremely well measured \cite{2020PhRvD.102l4037T, 2021PhRvD.104d4065B, 2023PhRvD.108b3022D, 2022arXiv220403423K}.
However, this complicates the initial search for candidate signals which have to be identified in a large astrophysical parameter space.

To address these challenges, several groups have tackled the SoBBH search problem.
Most follow a hierarchical and semi-coherent strategy; see, e.g.,\, \cite{2024arXiv240710797F}.
Semi-coherent methods split the signal into multiple segments (in either the time or frequency domain) before comparing it with a template.
Hierarchical searches typically begin by using a large number of semi-coherent segments to cover the space efficiently before refining to a smaller number to hone in on the true signal.
There are also machine learning approaches to the search problem \cite{2024arXiv240607336Z, 2022PhRvD.105l3027Z} (e.g.\ using convolutional neural networks) and archival searches \cite{2018PhRvL.121y1102W, 2021PhRvD.103b3025E} which wait until after the source merges in the LIGO/Virgo band before looking back at the LISA data.
This study uses a graphical processing unit (GPU) accelerated implementation of the hierarchical and semi-coherent search previously developed by the authors \cite{2023PhRvD.108h4014B, 2024arXiv240813170B}.

SoBBH searches have hitherto been run on simple datasets, specifically generated for the purpose of validating the search.
Here, results are presented for the first attempts at a SoBBH search on the multipurpose LISA Data Challenges (LDCs) the community is using to prepare for LISA.
Specifically, we use the 1b \Yorsh{} LDC \cite{LDC}, a dataset designed for developing methods for analyzing complex, long-duration signals.
This increases the difficulty because both the waveform and the LISA response model used in the search differ from those used to generate the data. 
Furthermore, because the search is performed by different people and codes from those that generated the data, it forces us to confront the practical problems of matching all the necessary conventions.
\Yorsh{} is not a \emph{blind} data challenge.

The search output is a point estimate for the parameters of a candidate signal that maximize the semi-coherent statistic $\Upsilon_{N=1}$ \cite{2024arXiv240813170B}.
The significance of the candidate is assessed using a background distribution of noise triggers.
Confident detections are followed up with rapid, low-cost parameter estimation using an ensemble Markov chain Monte Carlo (MCMC) \cite{karamanis2021zeus,2024arXiv240813170B}. 
This first estimate of the Bayesian posterior is intended to be used to initialize more detailed parameter estimation as part of a global fit \cite{2023PhRvD.107f3004L, 2024arXiv240504690K}.
Sec.~\ref{sec:results} contains the results of two versions of our search using different models for the instrument response and noise power spectral densities (PSDs).
The best performing search confidently identifies five sources, including one with an injected signal-to-noise ratio (SNR) as low as 12.94.
Rapid parameter estimation confirms these correspond to the five loudest injections with only small parameter biases expected due to differences in the waveform and response modeling.

In the rest of the paper,
Sec.~\ref{sec:data} describes the SoBBH part of the \Yorsh{} LDC.
Sec.~\ref{sec:methods} briefly describes the search, focusing on the differences from Ref.~\cite{2024arXiv240813170B}.
Sec.~\ref{sec:computation_cost} contains details of the computational costs of the search.
Sec.~\ref{sec:discussion} contains a discussion of the results.

\section{Data}\label{sec:data}

The data analyzed in this study was produced for the 1b \Yorsh{} LDC \cite{LDC}. The simulated data includes time series for the three second-generation time-delay interferometry (TDI-2) channels $(X,Y,Z)$, 2 years in duration and downsampled to a $5\,\mathrm{s}$ cadence. 
Our search uses all of the approximately noise-orthogonal $(A,E,T)$ TDI channels \cite{2002PhRvD..66l2002P} obtained by a linear combination of $(X,Y,Z)$.
Note that although the GW response in the $T$ channel is suppressed at low frequencies, and is consequently often neglected (see, e.g.\, Refs.~\cite{2023PhRvD.107f3004L, 2024arXiv240504690K}), it is more important at the high frequencies $\gtrsim 10\,\mathrm{mHz}$ where most SoBBH signals reside\footnote{
    For example, source \#8 has $\rho^{2}_A=276$, $\rho^{2}_E=292$ and $\rho^{2}_T=26$; the $T$ channel contains $\sim 4\%$ of the injected squared SNR. 
}.

\begin{figure}[t]
    \centering
    \includegraphics[width=0.5\textwidth]{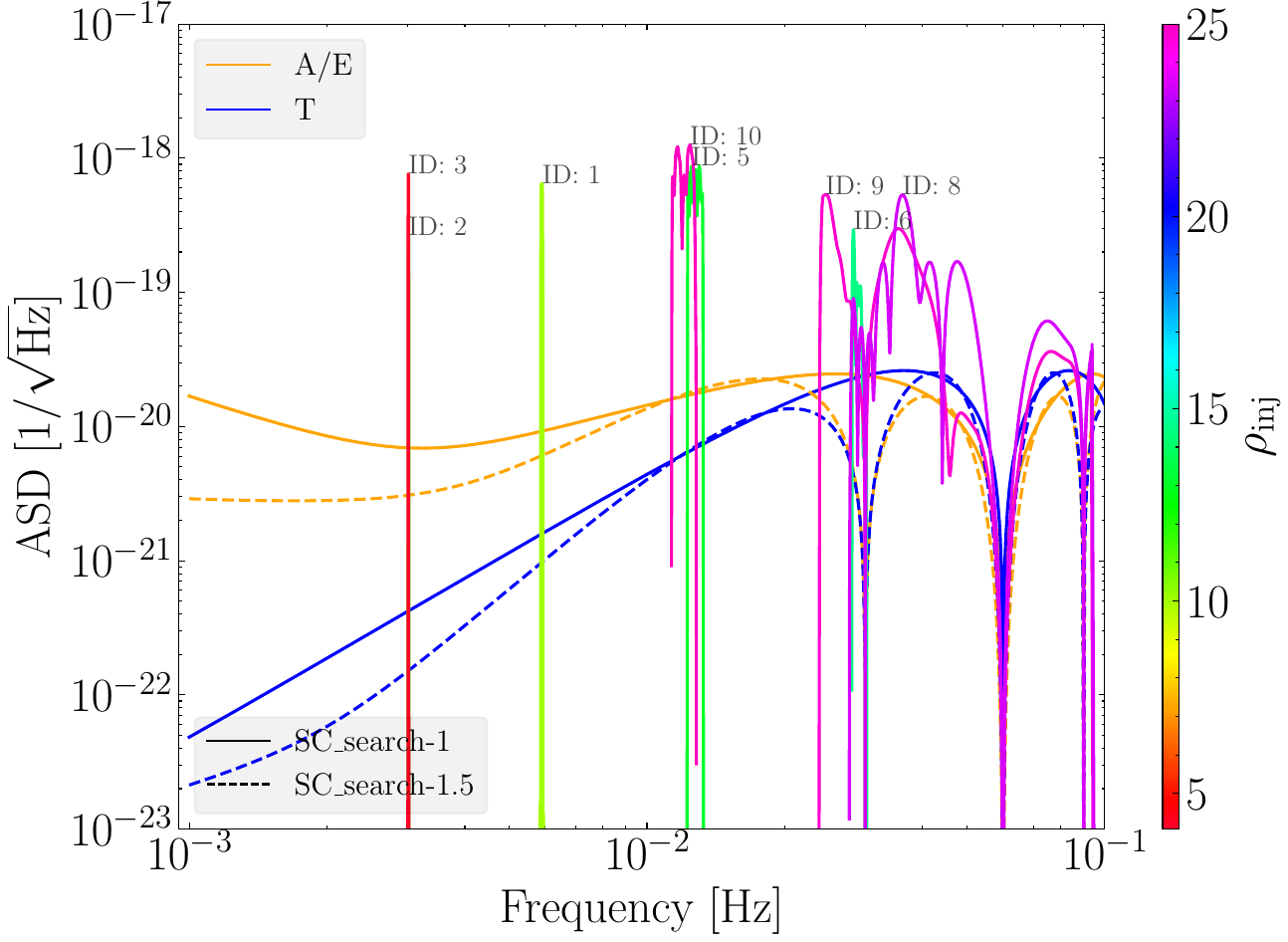}
    \caption{ \label{fig:data_and_PSDs}
    The high-frequency LISA bandwidth containing SoBBH signals. 
    Orange ($A$, $E$) and blue ($T$) curves show the ASDs in the TDI channels.
    Different line styles show the ASDs for the two searches.
    Note, the \texttt{SC\_search-1.5} ASD has three zeros in the range plotted while the \texttt{SC\_search-1} ASD has one. 
    Also plotted are representations of the signals for the sources in Table \ref{tab:injection_params} colored by injected SNR; the waveform quantity plotted is $2|\tilde{A}(f)|\sqrt{f}$. Sources \#2 and \#3 overlap on this plot and cannot be distinguished.
    }
\end{figure}

The simulated data includes idealized instrumental noise.
The noise model is stationary and Gaussian and was used to generate data in the frequency domain from a known PSD before transforming to the time domain. 
The PSD was the same as that used in the \Sangria{} LDC data challenge \cite{Sangria}, differing slightly from that in the LISA science requirements (SciRD) \cite{2021arXiv210801167B}. Hereafter, this will be referred to as the \Yorsh{} PSD.
Galactic confusion noise \cite{2021PhRvD.104d3019K,2021MNRAS.508..803B} was not included; while this will be a significant non-stationary noise source at lower frequencies $\lesssim 3\,\mathrm{mHz}$ \cite{2021PhRvD.104d3019K}, it is unlikely to significantly affect the search for the loudest SoBBH sources at high frequencies $\gtrsim 10\,\mathrm{mHz}$ (see Fig.~\ref{fig:data_and_PSDs}).


For the simulated data, the SoBBH injections were modeled using the quasi-circular, aligned-spin waveform \texttt{IMRPhenomD} \cite{2016PhRvD..93d4007K} (as implemented within \texttt{LISAbeta} \cite{lisabeta}) transformed into the time domain via the stationary phase approximation. 
The instrument response is modeled in the time domain, directly computing the relevant delays using \texttt{LISACode} \cite{2008PhRvD..77b3002P}.
There are eight SOBBH sources in \Yorsh{}, numbered 1 to 10 (4 and 7 missing).
Key parameters of the injections are summarized in Table \ref{tab:injection_params} and the signals are plotted in Fig.~\ref{fig:data_and_PSDs}.
These parameters cover a range of chirp mass, $\mathcal{M}_c$, time to merger $t_c$, and mass ratio, $q$.
These parameters were chosen to facilitate the development of data analysis algorithms and were not intended to be an astrophysically realistic population.
For example, sources \#2 and \#3 have $\mathcal{M}_c$ and $t_c$ values much closer together than would normally be expected to allow for investigations into source confusion.

All injections have component spins aligned with the orbital angular momentum and are therefore non-precessing.
LISA is not expected to be able to constrain the component spins of SoBBHs with high accuracy \cite{2020PhRvD.102l4037T, 2021PhRvD.104d4065B, 2023PhRvD.108b3022D, 2022arXiv220403423K} but selected results are presented in Sec.~\ref{sec:results} for the effective aligned spin combination $\chi_{\rm eff}$.

Table \ref{tab:injection_params} gives both the time to merger, $t_c$, and the GW frequency at the start of observations, $f_{\rm low}$. 
These are not independent; the search works with $f_{\rm low}$ directly and $t_c$ is a derived parameter, computed using \texttt{TaylorF2Ecc}.

Table \ref{tab:injection_params} also reports the injected SNRs of the sources. 
These were calculated by taking a discrete Fourier transform of the injected signal after applying a Tukey window with parameter $\alpha=0.1$ \cite{github_repo}. 
The same window was used to prepare the frequency domain data for the search.

\section{Methods}\label{sec:methods}

The semi-coherent search algorithm (\texttt{SC\_search}) used here closely follows that described in Ref.~\cite{2024arXiv240813170B}.

The search uses a custom implementation of the \texttt{TaylorF2Ecc} waveform model \cite{2016PhRvD..93l4061M}.
Note that this differs from the waveform used in the injections.
This frequency-domain, $\ell=|m|=2$ mode, post-Newtonian (PN) waveform approximant includes contributions to the GW phase from small orbital eccentricities, vital for the SoBBH population that LISA is expected to see \cite{2024PhRvD.110f3012F}.
Note that even though the injections in \Yorsh{} were performed with quasi-circular waveforms, the search does not assume this and includes an additional parameter, $e_0$, for the binary eccentricity at the start of LISA observations.
This waveform includes eccentric phase contributions up to 3 PN order expanded to leading order in $e_0$.
Compared to what was used in Ref.~\cite{2024arXiv240813170B}, the waveform has been further extended to include leading-order contributions to the GW phase due to the aligned component spins. 
The corrections to the phase arising from the spin contribution can be found in Ref.~\cite{2022PhRvD.105b3003F} and references therein. 
All aligned spin terms up to 2.5 PN order were included. The new spin terms have been checked in the quasicircular $e_0=0$ limit by comparing with the \texttt{TaylorF2} implementation used in Ref.~\cite{2021PhRvD.104d4065B}.

The search uses a fast model of the LISA response to convert the GW waveform, represented in the frequency domain using amplitude and phase, to time-delay interferometry (TDI) variables $(A,E,T)$.
Currently, there does not exist a fast GPU implementation of the TDI-2 response suitable for use in this sort of analysis.
Therefore, two other models of the instrument response were used.
Both searches use the frequency-domain response described in Refs.~\cite{2021PhRvD.103h3011M,2018arXiv180610734M} and implemented in \texttt{BBHx} \cite{2020PhRvD.102b3033K}. 
\texttt{SC\_search-1} used the first-generation TDI-1 variables $(A,E,T)$ that assume a static LISA constellation. 
\texttt{SC\_search-1.5} used the improved TDI-1.5 variables $(A,E,T)$, which allow for a rigid (i.e.,\ constant armlength) rotation of the constellation.
Neither used the TDI-2 variables used in the injections that allow for relative motion between the spacecraft \cite{2022PhRvD.105f2006H, 2021PhRvD.104b3006B}.

The search assumes a known noise PSD, $S_{\alpha}(f)$ for $\alpha\in\{A,E,T\}$.
\texttt{SC\_search-1} used the SciRD model for the instrumental noise in the TDI-1 channels, from Ref.~\cite{2021arXiv210801167B}. 
\texttt{SC\_search-1.5} used the LDC \Yorsh{} 
PSD for the TDI-2 channels (see Sec.~\ref{sec:data}).
Double white dwarf confusion was not included as \Yorsh{} does not include these sources.
These PSDs are plotted as amplitude spectral densities (ASDs, $\sqrt{S_\alpha(f)}$) in Fig.~\ref{fig:data_and_PSDs}. 

\begin{figure}[t]
    \centering
    \includegraphics[width=0.49\textwidth]{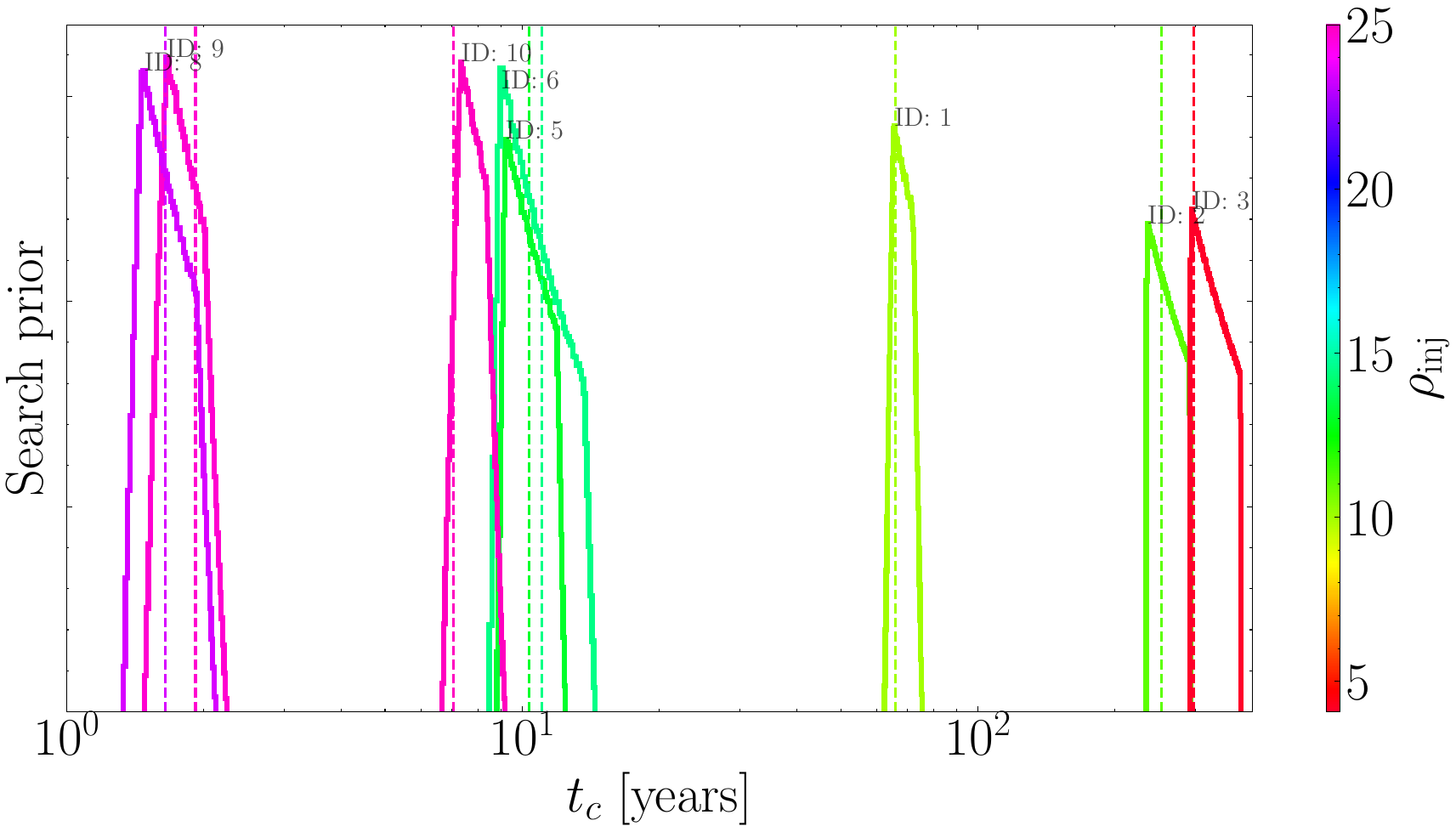}
    \caption{ \label{fig:time_to_merger}
    One-dimensional marginalized prior probability on the time to merger parameter $t_c$ used in the search. 
    Flat priors are used for the $\mathcal{M}_c$, $f_{\rm low}$ parameters within the restricted ranges described in the main text. 
    When transformed to $t_c$, these priors give the distributions shown here.
    The injected values of $t_c$ are shown for each source using the vertical dashed lines.
    Colors for all eight sources match those in Fig.~\ref{fig:data_and_PSDs}.}
\end{figure}

As described in Ref.~\cite{2024arXiv240813170B}, the search does not cover the entire SoBBH parameter space at once.
Instead, restricted search priors were used for $\mathcal{M}_c$ and $f_{\rm low}$; this part of parameter space is referred to as a search tile.
The two-dimensional $\mathcal{M}_c$, $f_{\rm low}$ space can then be covered by $\approx 100-1000$ search tiles, which run independently and in parallel.
In this study, to limit the computational cost, one separate search tile was used for each source in Table \ref{tab:injection_params}. 
The process of choosing the tile boundaries has not yet been automated and was therefore done by hand.
The width of each tile in $\mathcal{M}_c$ was set to approximately $\sim 5 \rm{M_{\odot}}$.
The width of each tile in $f_{\rm{low}}$ was chosen such that the prior width on the derived parameter $t_c$ was approximately equal to the actual time to merger of that source.
Figure \ref{fig:time_to_merger} shows the induced search priors on the derived parameter $t_c$ for all eight sources.
The exact values for the tile boundaries are given in Ref.~\cite{github_repo}.

Within each search tile, both the search and the subsequent rapid parameter estimation (if performed) use flat priors in all of the parameters described in Table I of Ref.~\cite{2024arXiv240813170B}. 
This includes the parameters $\mathcal{M}_c$ and $f_{\rm low}$ used to define the tile boundaries.
Additionally, flat priors were used on the individual component spin magnitudes in the range $[-0.99, 0.99]$.
Together with the flat priors on the chirp mass and symmetric mass ratio, this gives the non-flat prior on the $\chi_{\rm eff}$ parameter shown in the bottom right panel of Fig.~\ref{fig:post}.

Each search attempts to find the loudest candidate in its search tile.
The statistical significance of these candidates is assessed by comparing the value of the $\Upsilon_{N=1}$ statistic against a background noise distribution.
The background should be generated by running a large number of identical searches on simulated LISA datasets which do not contain SoBBHs.
For more details, see the demonstration of this process in Ref.~\cite{2024arXiv240813170B}.
This process is computationally expensive and must be repeated for each tile to account for variations in the background distribution across parameter space. 
Therefore, the background noise distribution from Fig.~5 in Ref.~\cite{2024arXiv240813170B} was used for all tiles in this study.
Here, sources with $\Upsilon_{N=1}\geq 100$ are deemed to be confidently detected.

Confident detections were followed up with rapid parameter estimation.
The rapid parameter estimation settings were identical to those in Ref.~\cite{2024arXiv240813170B}. 
The ensemble MCMC used $10^2$ walkers, with $10^4$ iterations.

\section{Results}\label{sec:results}

The main results are summarized in Table \ref{tab:search_results}, which reports the results of the searches, and Table \ref{tab:rapid_PE_results}, which reports the results of the rapid parameter estimation.  

\texttt{SC\_search-1} only finds two of the eight sources.
This is partially expected given that it uses an inferior (lower-generation) TDI than that used in the injections.
It is interesting that this search found source \#5 with injected SNR $12.94$ while missing the louder sources \#6, \#8 and \#9.
This is likely because source \#5 was at lower frequency and where the differences between the TDI generations is less pronounced (see, e.g.,\, the ASD curves in Fig.~\ref{fig:data_and_PSDs}).

%
%
\begin{table*}
    \caption{ \label{tab:injection_params}
    \textbf{Injection parameters}. 
    Source parameters used to produce the injected signals in the \Yorsh{} LDC.
    The final column reports the optimal SNR of each injection ($\rho^{\rm inj}$) computed using the injected waveforms (from the noiseless datastreams) and the LDC-2 PSD.
    All of the injected binaries were on quasi circular orbits, i.e.\ with zero injected eccentricity. Events marked with a star* are those that merge within the LISA mission duration and are often described as multiband sources. 
}
\begin{ruledtabular}
\begin{tabular}{|l|cccccc|c|}
    \textrm{ID} &
    $\mathcal{M}_c^{\rm inj} \,  [M_{\odot}]$ &
    $t_c^{\rm inj} \, [\rm{years}]$ & 
    $f_{\rm low}^{\rm inj}\, [\rm{mHz}]$ & 
    $d_L^{\rm inj} \, [\rm{Mpc}]$ & $q^{\rm inj}$ & $\chi_{\rm eff}^{\rm inj}$ &
    $\rho^{\rm inj}$\\
    \hline \hline 
    \#1 & $29.34741587$ & $\phantom{0}65.9176$ & $\phantom{0}5.85830665$ & $159.9$ & $0.91$ & $\phantom{-}0.50\phantom{0}$& $10.91$ \\
    \#2 & $38.04622881$ & $252.7789$ & $\phantom{0}3.00851783$ & $\phantom{0}94.5$ &  $0.83$ & $-0.06\phantom{0}$ & $4.07$ \\
    \#3 & $34.51216704$ & $297.7712$ & $\phantom{0}3.00698596$ & $\phantom{0}47.0$ &  $0.58$ & $\phantom{-}0.10\phantom{0}$ & $9.88$ \\
    \#5 & $27.41970433$ & $\phantom{0}10.3457$ & $12.24273032$ & $168.3$ &  $0.83$ & $-0.55\phantom{0}$ & $12.94$ \\
    \#6 & $7.007404972$ & $\phantom{0}11.0420$ & $28.02352272$ & $\phantom{0}17.3$ &  $0.88$ & $-0.17\phantom{0}$ & $14.30$ \\
    \#8* & $22.40969304$ & $\phantom{00}1.6501$ & $27.65438527$ & $\phantom{0}34.0$ &  $0.59$ & $\phantom{-}0.002$ & $24.37$ \\
    \#9* & $26.08583360$ & $\phantom{00}1.9185$ & $23.76783772$ & $\phantom{0}85.5$ &  $0.95$ & $\phantom{-}0.10\phantom{0}$ & $23.08$\\
    \#10 & $39.14942200$ & $\phantom{00}7.0604$ & $11.31112717$ & $168.9$ &  $0.88$ & $\phantom{-}0.03\phantom{0}$ & $24.65$ 
\end{tabular}
\end{ruledtabular}
    %
    %
\caption{ \label{tab:search_results}
    \textbf{Search results}. 
    Two searches were performed for each source using the different models for the LISA response and different PSDs described in the main text. 
    Results from both searches are reported here. 
    Sources that were confidently detected are indicated with a tick and their chirp mass and time to merger parameter values are reported as differences from the injected values; e.g.,\ $\delta \mathcal{M}_c = \mathcal{M}_c^{\rm search}-\mathcal{M}_c^{\rm inj}$.
    The matched filter SNR, $\rho_{\rm mf}$, of the template that maximized the semi-coherent detection statistic is also reported.
    For sources that were not confidently detected, those indicated by a cross, the search did not find the injected source; i.e.\ the maximum-search-statistic-parameters did not correspond to the injection. Therefore, these non-detections represent sources which were missed by the search, not sources that were found and assigned a low significance.
}
\begin{ruledtabular}
\begin{tabular}{|l|cccc|cccc|}
    &\multicolumn{4}{c|}{\texttt{SC\_search-1}} & 
    \multicolumn{4}{c|}{\texttt{SC\_search-1.5}} \\
    \cline{2-9}
    \textrm{ID} &
    \textrm{Found} & 
    $\delta\mathcal{M}_c\, [M_{\odot}]$ &
    $\delta t_c \, [s] $ & $\rho_{\rm mf}$ &
    \textrm{Found} & 
    $\delta \mathcal{M}_c, [M_{\odot}]$ &
    $\delta t_c [s]$ & $\rho_{\rm mf}$ \\
    \hline \hline
    \#1 & \crossmark & - & - & - & \crossmark & - & - & - \\ 
    \#2 & \crossmark & - & - & - & \crossmark & - & - & - \\ 
    \#3 & \crossmark & - & - & - & \crossmark & - & - & - \\ 
    \#5 & \checkmark & $0.0015$ & $42482$ & $12.2$ & \checkmark & $0.0004$ & $40689.$ & $11.60$  \\ 
    \#6 & \crossmark & - & - & - & \checkmark & $0.0005$ & $28031.$ & $14.91$ \\ 
    \#8 & \crossmark & - & - & - & \checkmark & $0.0001$ & $611.$ & $21.33$ \\ 
    \#9 & \crossmark & - & - & - & \checkmark  & $0.0020$ & $1395.$ & $23.53$ \\ 
    \#10 & \checkmark & $0.0002$ & $24297$ & $26.54$ & \checkmark & $0.0006$ & $22137.$ & $25.77$
\end{tabular}
\end{ruledtabular}
    %
    %
\caption{ \label{tab:rapid_PE_results}
    \textbf{Rapid parameter estimation results}. 
    Here we report the results of the rapid parameter estimation performed by the simple MCMC algorithm which can be performed automatically at the end of a search when a source is found.
    For searches that did not find a candidate in Table \ref{tab:search_results}, the corresponding cell in the table is left blank.
    For the chirp mass and time to merger, the values reported in the table are the widths (90\% credible interval) of the posteriors which are all centered near the search values reported in Table \ref{tab:search_results}.
    For the luminosity distance and the eccentricity, the median and 90\% credible intervals are reported in the form of an asymmetric error bar.
    The (90\% credible) sky area for each source, $\Omega_{90}$, is also reported.}
\begin{ruledtabular}
\begin{tabular}{|l|ccccc|ccccc|}
    &\multicolumn{5}{c|}{PE from \texttt{SC\_search-1}} & 
    \multicolumn{5}{c|}{PE from \texttt{SC\_search-1.5}} \\
    \cline{2-11}
    \textrm{ID} &
    $\Delta\mathcal{M}_c\,[M_\odot]$ & $\Delta t_c$ [hr] & $d_L\,[\mathrm{Mpc}]$ & $\Omega_{90}$ [deg$^2$] & $e_0 \, [\times 10^{-3}]$ &
    $\Delta\mathcal{M}_c\,[M_\odot]$ & $\Delta t_c$ [hr] & $d_L\,[\mathrm{Mpc}]$ & $\Omega_{90}$ [deg$^2$] & $e_0\, [\times 10^{-3}]$\\
    \hline \hline
    \#1 & - & - & - & - & - & - & - & - & - & -\\ 
    \#2 & - & - & - & - & - & - & - & - & - & -\\  
    \#3 & - & - & - & - & - & - & - & - & - & - \\  
    \#5 & $0.01$ & $21.80$ & $121_{-42}^{+30}$ & $25.42$  & $5.1_{-5.1}^{+4.3}$ & $0.01$ & $21.63$ & $237_{-86}^{+61}$  & $25.87$ & $5.4_{-5.4}^{+4.0}$ \\ 
    \#6 & - & - & - & - & - & $0.002$ & $19.22$ & $17_{-4}^{+15}$ & $0.65$ & $6.0_{-3.6}^{+3.1}$ \\ 
    \#8* & - & - & - & - & - & $0.001$ & $0.19$ & $39_{-3}^{+4}$ & $11.51$ & $3.3_{-0.6}^{+1.1}$ \\ 
    \#9* & - & - & - & - & - & $0.001$ & $0.07$ & $77_{-8}^{+12}$ & $1.16$ & $3.2_{-0.5}^{+0.4}$  \\ 
    \#10 & $0.01$ & $9.7$ & $84_{-26}^{+15}$ & $1.05$ & $4.8_{-4.8}^{+2.8}$ & $0.01$ & $10.32$ & $156_{-48}^{+34}$ & $1.98$ & $4.7_{-4.7}^{+2.9}$
\end{tabular}
\end{ruledtabular}
\end{table*}

\texttt{SC\_search-1.5} finds all five sources in the data challenge with injected signal-to-noise ratios $\gtrsim 12$. 
This is impressive and suggests a threshold SNR necessary for the detection of these SoBBH sources significantly lower than some previous estimates \cite{2024arXiv240813170B,2019MNRAS.488L..94M,2024arXiv240607336Z}.
The expected number of SoBBH signals depends strongly on the threshold SNR, a lower threshold implying a higher source count. 
The threshold SNR will not be accurately known until a search pipeline has been fully developed and run on realistic LISA data (e.g.,\ including realistic noise, gaps, and other sources) to estimate the background distribution of noise triggers.
This is complicated by the fact that the threshold SNR will vary significantly across the large SoBBH parameter space.

The \texttt{SC\_search} results are summarized in Table \ref{tab:search_results}.
In cases where the maximum of the semi-coherent statistic $\Upsilon_{N=1}$ did not exceed the threshold, no confident candidate source was found;
these are marked with a cross in Table \ref{tab:search_results} and no parameters are reported.
When a source is not found this is typically because the source is too quiet or, in the case of \texttt{SC\_search-1}, because a loud source (such as \#6, \#8 and \#9)  accumulates significant SNR at frequencies around $\sim 25\,\mathrm{mHz}$ where the simplified TDI-1 response used by \texttt{SC\_search-1} gives a poor approximation to TDI-2 used for the injections and it is not possible to match the template signal.
When a source is found, the parameters of the template that maximizes the detection statistic are reported in the table.

It is interesting to note that both searches made detections using TDI generations that differ from the injections. 
This shows that for some sources, particularly at lower frequencies, it is possible to perform LISA data analysis without the most accurate model for the instrument response.
This will likely be true for other source types and may allow for the use of simpler and faster response models in carefully selected parts of a global fit.

Rapid parameter estimation was performed for all of the confident detections produced by both searches. 
The parameter estimation used the same model for the instrument response as the initial search.
The rapid parameter estimation provides further confirmation that the high-confidence detections produced by the search do correspond to the injected sources in the data challenge. 
The posteriors are extremely narrow in several key parameters (notably, $\mathcal{M}_c$ and $f_{\rm low}$) and are close to the injected values. An example posterior is shown in Appendix \ref{app:posteriors}.

Several parameter biases are observed. This is expected due to the use of different waveform and instrument response models for the injection and recovery. The largest biases are in some of the distance parameters in \texttt{SC\_search-1}; see, e.g.\ the source \#10 in Appendix \ref{app:posteriors}. The factor by which the distance is underestimated can be understood from the definitions of the TDI-1 variables which are related to the TDI-1.5 variables by a factor $(1-\exp(-8\pi i fL))$ where $L$ is the LISA armlength (see, for example, the implementation in \texttt{BBHx} \cite{BBHx_git}).

Other small biases are observed, such as the $e_0$ posteriors which occasionally peak at non-zero values. Also, some $\chi_{\rm eff}$ posteriors (notably sources \#8 and \#9) peak away from the injected values. These are likely due to the different waveform used for the injection and recovery.

As expected, LISA SoBBH observations are not able to significantly constrain the component spins. 
An example of the posterior on $\chi_{\rm eff}$ for source \#10 is shown in Appendix \ref{app:posteriors}; although low values of $\chi_{\rm eff}$ are preferred, no parameter range is confidently excluded.
LISA measurements also allow the time of merger to be predicted within a few hours.
This is consistent with previous parameter estimation studies (see, e.g.\ Refs.~\cite{2020PhRvD.102l4037T, 2021PhRvD.104d4065B, 2023PhRvD.108b3022D, 2022arXiv220403423K}) .

Table \ref{tab:rapid_PE_results} also reports results for sky localizations obtained from the rapid parameter estimation analyses. As long-lived and high-frequency LISA sources, SoBBHs are typically quite well localized; all the recovered sky maps contain a single, roughly Gaussian posterior mode.
The areas were calculated using a 2-dimensional kernel density estimate (KDE) of the posterior probability for the ecliptic longitude and sine latitude, $(\lambda, \sin\beta)$, obtained by performing a Monte Carlo integral with samples drawn from the KDE.

\section{Computational Cost}\label{sec:computation_cost}

The computational costs of the \texttt{SC\_search} are similar to those described in Ref.~\cite{2024arXiv240813170B}. Each tile used 1 NVIDIA A100 GPU and 10 CPUs and took $\mathcal{O}(1-3)$ days, with the highest frequency searches taking the longest. The searches for different tiles were run in parallel. These costs may be further reduced with methodological improvements; therefore, this estimate should be treated as an upper bound for the cost of the SoBBH search.

\section{Discussion}\label{sec:discussion}

This paper presented results from the first SoBBH searches in the LDCs.
The most effective search successfully identifies all sources with injected SNRs $\gtrsim 12$. 
This is an encouraging result, as well as being somewhat unexpected given previous estimates for the threshold SNR. The search also automates rapid parameter estimation and was able to provide crude estimates of the posterior distributions suitable for initializing more detailed parameter estimation, e.g.\ as part of a global fit.

The main limitation of the \texttt{SC\_search} currently is the lack of a suitable fast implementation of the LISA response.
This should be developed as a priority and will be incorporated into this search as soon as it is available.

The other limitations concern the idealized nature of the LDC data. 
The \Yorsh{} data was generated without gaps or glitches, without other source classes (such as high-frequency double white dwarfs or extreme-mass-ratio inspirals), without uncertainty in the instrumental noise, and without non-stationary or non-Gaussian noise components.
Subsequent data challenges have already started to relax some of these assumptions but so far do not include SoBBHs. 
This search will continue to be applied to future LDCs containing SoBBHs.

\begin{acknowledgments}
    We thank the UK LISA Ground Segment team and the LISA DDPC CU-L2D for useful discussions. We thank Stanislav Babak, Francisco Duque and Alvin Chua for helpful comments on the manuscript.
    CM acknowledges support from ST/Y004922/1, ST/V002813/1 and ST/X002071/1. DB is supported by the STFC.
    Computations were performed using the University of Birmingham’s BlueBEAR HPC service. 
    DB acknowledges the support of the Google Cloud Research Credits program, Grant No.\ 289387648. 
    This paper made use of CuPy, Numba, NumPy and Matplotlib \cite{Okuta:2017, Lam:2015,numpy,matplotlib}, 
    and the following LISA packages: \texttt{BBHx} \cite{michael_katz_2023_7791640}, \texttt{LISAbeta} \cite{lisabeta}, and \texttt{LISACode} \cite{2008PhRvD..77b3002P}.
\end{acknowledgments}

\section*{Code availability}
The \texttt{SC\_search} code is available at Ref.~\cite{search_git}.
Results and data products are available at Ref.~\cite{github_repo}.


\bibliographystyle{apsrev4-1}
\bibliography{bibliography}


\onecolumngrid
\appendix

\section{Posterior plots} \label{app:posteriors}

\begin{figure*}[b]
    \includegraphics[width=0.7725\textwidth]{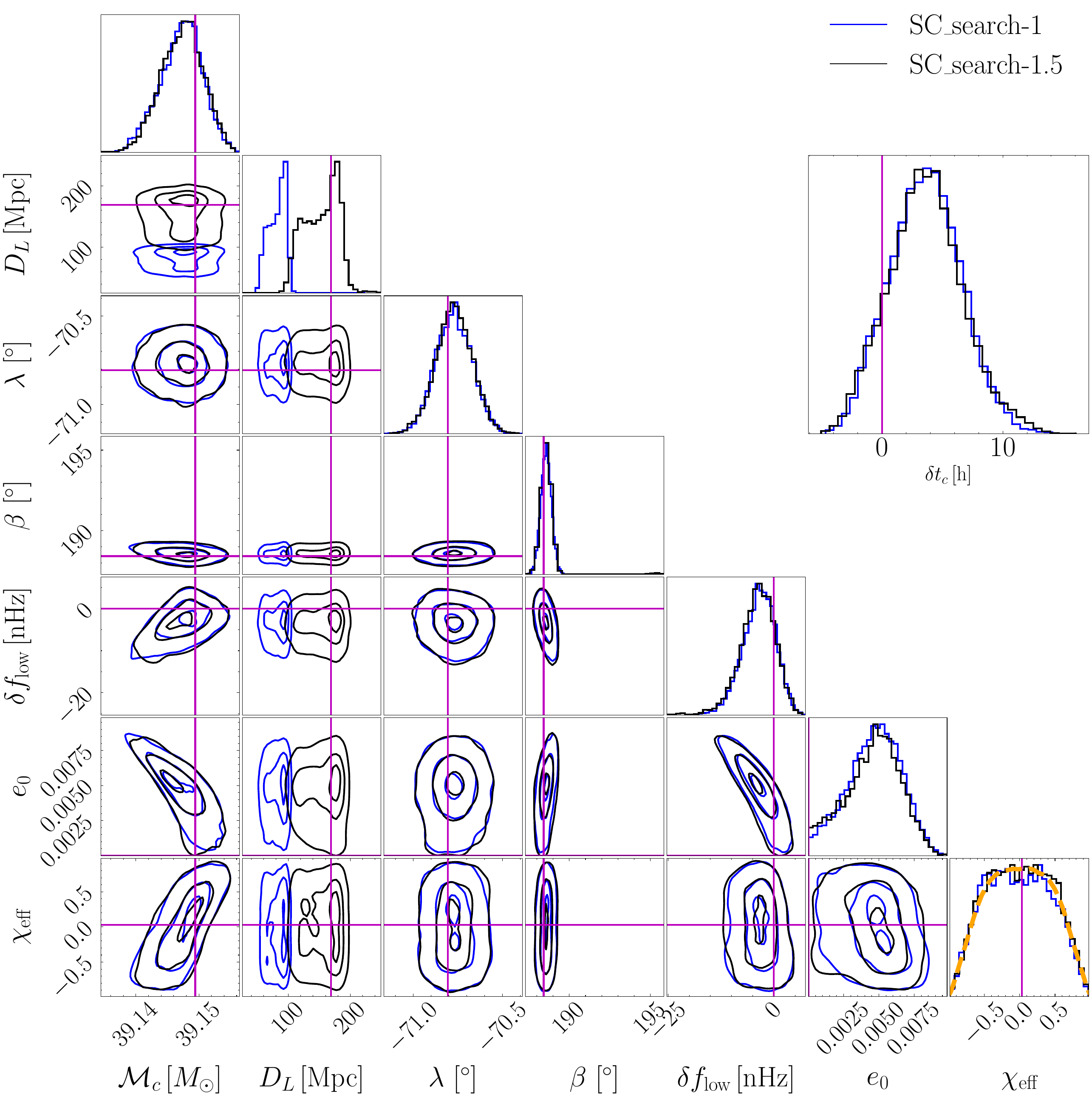}
    \caption{\label{fig:post}
        Rapid parameter estimation posteriors on selected parameters for source \#10.
        Magenta lines indicate the injected parameters.
        Two-dimensional contours show the 10, 50, and 90\% credible regions.
        The \texttt{SC\_search-1} posterior significantly underestimates the distance to the source due to the use of a lower generation TDI, as explained in the text.
        The \texttt{SC\_search-1.5} posterior is consistent with all the injected parameters.
        Also shown in the top right is the posterior on the derived parameter $\delta t_c = t_c-t_c^{\rm inj}$. 
        The orange curve shows the 1-dimensional marginalized prior distribution on the $\chi_{\rm eff}$ parameter.
    }
\end{figure*}

Included here is an example of the posteriors produced by the rapid parameter estimation performed at the end of the search. 
Fig.~\ref{fig:post} shows the posterior for selected parameters of the loudest source \#10.
Posteriors for other sources can be found in Ref.~\cite{github_repo}.
We stress that these posteriors are produced by rapid parameter estimation.
The emphasis here was on keeping the computational costs down and no detailed convergence checks were performed. It is intended that these are used to initialize subsequent detailed parameter estimation, e.g.\ as part of a global fit.

\end{document}